\documentclass[%
 reprint,
amsmath,amssymb,
aps,
pre,
]{revtex4-2}

\usepackage{graphicx}
\usepackage{dcolumn}
\usepackage{bm}
\usepackage{braket}

\usepackage{algpseudocode,algorithm}
\makeatletter
\renewcommand{\ALG@name}{}

\makeatother

\usepackage{color}
\usepackage{colortbl}

\begin{document}

\preprint{APS/123-QED}

\title{Transfer learning for nonlinear dynamics and its application to fluid turbulence}

\author{Masanobu Inubushi}
 \email{inubushi@me.es.osaka-u.ac.jp}
\author{Susumu Goto}%
\affiliation{%
Graduate School of Engineering Science,
Osaka University,\\
1-3 Machikaneyama, Toyonaka,
Osaka 560-8531, Japan.}%




\date{\today}

\begin{abstract}
We introduce transfer learning for nonlinear dynamics, which enables efficient predictions of chaotic dynamics by utilizing a small amount of data.
For the Lorenz chaos,
by optimizing the transfer rate, we accomplish more accurate inference than the conventional
method by an order of magnitude.
Moreover,
a surprisingly small amount of learning is enough to infer the energy dissipation rate
of the Navier-Stokes turbulence
because we can, thanks to the small-scale universality of turbulence,
transfer a large amount of the knowledge learned from turbulence data at lower Reynolds number.
\end{abstract}

\maketitle


\section{Introduction}
Machine learning (ML) is becoming a powerful tool for a broad range of problems in physics,
and it is likely to solve certain long-standing problems in nonlinear physics, such as turbulence modeling \cite{turbulencemodel, PhysRevFluids.2.054604, Fukagata2019}, in the near future.
Solving these problems is not only crucial in fundamental physics,
but also has an immeasurable impact upon practical problems,
e.g., in fields such as
mechanical engineering and weather forecasting.

As an ML method suitable for nonlinear dynamics, we focus on {\it reservoir computing (RC)} \cite{rcbook}. 
RC has been successfully applied to the problems of nonlinear dynamics
such as the inference of unmeasured variables
and the prediction of future states of
spatiotemporal chaos \cite{Lu2017, Parlitz2018, nakai2018, Soriano2019, Pathak2018, IY2017}.
Similarly to other ML methods, RC requires a large amount of training data.
However, this requirement is often unsatisfied,
i.e., the amount of training data is often limited.
This fundamental problem would be a major bottleneck of making RC applicable,
especially for spatiotemporal chaos with a large degree of freedom.

The ML-based turbulence modeling is a typical example.
Although the ML-based turbulence modeling will be useful in practical engineering applications,
these modeling requires a large amount of high-Reynolds-number turbulence data collected over a long period of time.
The length of the training data required for RC
exceeds the turnover time of the largest eddies by several thousand times,
for which we give a theoretical estimation in Appendix \ref{data_length}.
Generating such turbulence data for practical applications, e.g., using the direct numerical simulation, is usually impossible.
Therefore, it is essential to learn the knowledge of the high-Reynolds-number turbulence from a small amount of data.

To solve the fundamental problem, we employ {\it transfer learning},
which is a concept to utilize
knowledge learned in a task for another different but similar task.
Although transfer learning has been successfully used for ML tasks such as image classifications \cite{Weiss2016}, 
it is unclear how useful this concept is and how to implement it for problems in physics.

In this paper, we develop a transfer learning method for RC \cite{inubushi2019} with an optimization of {\it transfer rate} (defined below),
and then, we show that the method is indeed effective for tasks of chaotic dynamics.
Taking the Lorenz equations as an example,
we demonstrate that optimizing the transfer rate is essential,
which leads to
more accurate inference than the conventional method by an order of magnitude.

More importantly,
concerning the inference of the energy dissipation rate of the Navier-Stokes turbulence,
we uncover that
the amount of learning can be drastically reduced,
by transferring the knowledge learned from turbulence data at lower Reynolds number.
A main conclusion is that
the universality of the energy cascade of turbulence enables us to use a large transfer rate,
which is crucial for the ML-based turbulence modeling.



%

\section{Method}
We study a dynamical system ${\bm x}({k+1}) = {\bm f}_\rho ({\bm x}(k))$ with some control parameter $\rho$
and the inference task,
although our proposed method is not limited to this task.
The goal of the task is to infer an unmeasurable quantity $v({\bm x}(k))$ from a measurable quantity $u({\bm x}(k))$
at a parameter denoted by $\rho=\rho_{\mathcal{T}}$.
The training data $\mathcal{D_T}$ consists of the input data $u(k) = u({\bm x}(k))$ and the output data $v(k) = v({\bm x}(k))$,
i.e., $\mathcal{D_T}= \{ u(k), v(k) \}_{1 \le k \le T_L'}$.
We consider that the length of the training data $T_L'$ is not sufficiently long.

Here we assume that a sufficient amount of training data $\mathcal{D_S}= \{ u(k), v(k) \}_{1 \le k \le T_L}$ is available at a parameter $\rho=\rho_{\mathcal{S}}$ which differs from the parameter $\rho_{\mathcal{T}}$,
and these training data $\mathcal{D_T}$ and $\mathcal{D_S}$ are similar to each other.
We refer to the parameter $\rho_{\mathcal S}$ and $\rho_{\mathcal T}$ as the {\it source} and {\it target} domains (parameters), respectively.
Our method utilizes knowledge learned from the source domain to realize the inference in the target domain (see Fig.~1).

To derive explicit formulas,
we use the echo state network (ESN) introduced by Jaeger (2001)\cite{jaeger2001echo}
as a conventional RC method.
The state variable $r_i$ of the $i$-th node in the ESN evolves in time as follows:
$r_i(k+1) = \phi \Big[ \sum_{j=1}^N J_{ij} r_j (k) + \epsilon u(k) + \eta \xi_i \Big]$,
where $N$ is the number of nodes, and $J_{ij}$ and $\xi_{i}$ are the fixed random connections and biases, respectively.
Further, $\phi[ \cdot ]$ is the so-called activation function, and we employ $\phi[x] = \tanh [gx]$.
Here $g, \epsilon, \eta \in \mathbb{R}$ are hyper-parameters.
We used $N=100$ nodes of ESN with the following hyper-parameters: $g=0.95, \epsilon=0.2$, and $\eta=0.01$.
Elements $J_{ij}$ of the connection matrix and the bias terms $\xi_i$
are independently and identically drawn from the Gaussian distribution: $J_{ij} \sim \mathcal{N}(0, 1/N)$ and $\xi_i \sim \mathcal{N}(0, 1)$.
The time-series $\{ u(k), v(k) \}$ are normalized so as to have zero mean and unit variance.

\begin{figure}[t]
\includegraphics[width=8.6cm]{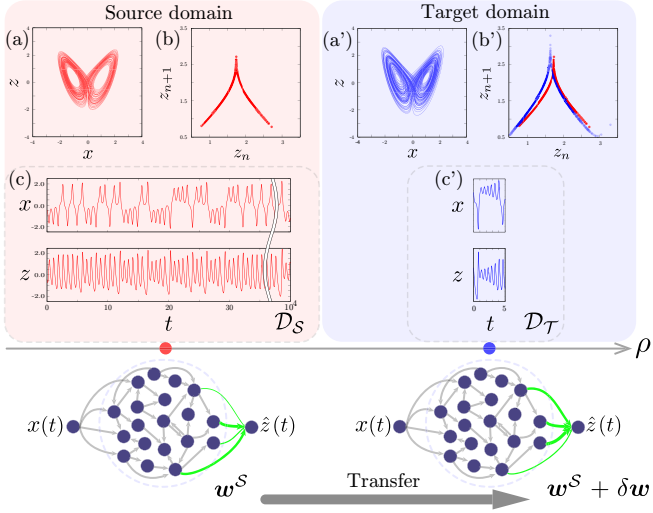}
\caption{Illustrative example of transfer learning for RC applied to the inference problem of $z(t)$ from $x(t)$ of the Lorenz chaos.
The red-shaded region on the left shows the data of the Lorenz chaos
in the source domain ($\rho_{\mathcal{S}} = 28$).
The blue-shaded region on the right shows the data in the target domain ($\rho_{\mathcal{T}} = 40$).
(a, a') Projection of the attractor onto the $x-z$ plane.
(b, b') Lorenz plot. The data plotted in (b) is replotted in (b') for comparison.
(c, c') Training data in $\mathcal{D_S}$ and $\mathcal{D_T}$. We consider the case that the training data in $\mathcal{D_T}$ is limited.
}
\label{concept_TL}
\end{figure}

\begin{figure*}[t]
\includegraphics[width=17.2cm]{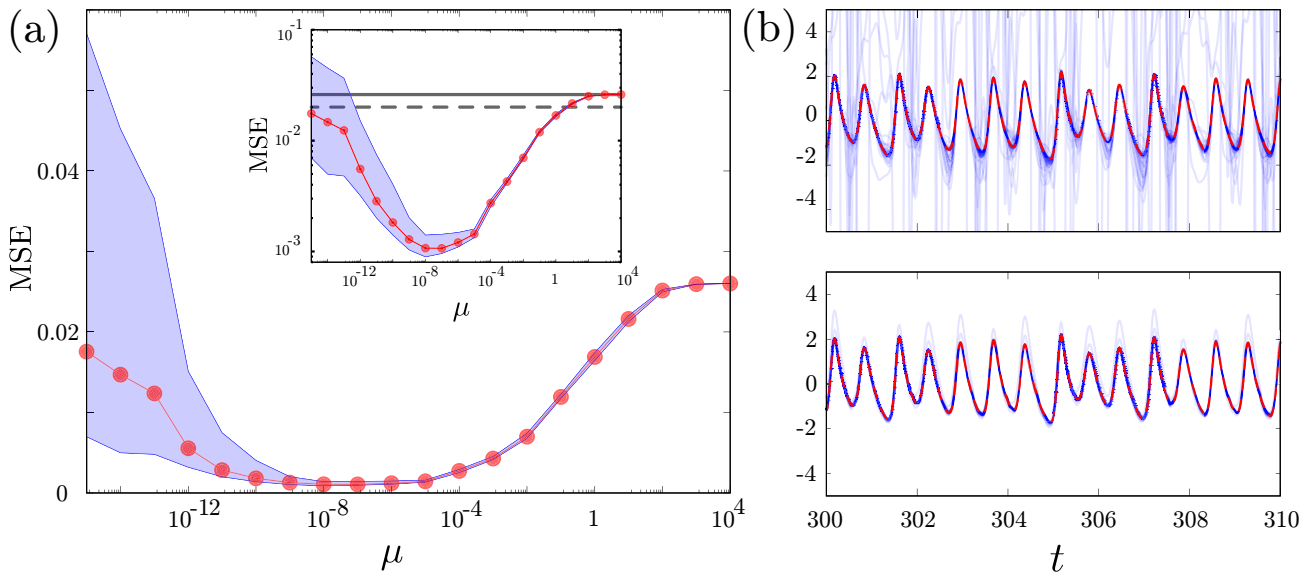}
\caption{(a)  Generalization error
(mean square error, MSE) of the inference task for the Lorenz equations with $ \rho_{\mathcal{T}}=32$.
The horizontal axis shows the transfer rate $\mu$. The red circles represent
the median MSE.
The blue shaded area indicates the range of the MSEs from the first to the third quartile.
Inset: the logarithmic graph of the same MSE data.
The broken and solid lines represent the median MSE in the cases of $\mu=0$ and $\mu=\infty$, respectively.
(b) The blue lines are the inferred values $\hat{z}^{(j)}(t)~~ (j=1,\cdots, 100)$
by the conventional method ($\mu=0$, upper panel) and the proposed transfer learning ($\mu=10^{-8}$, lower panel).
The red broken lines represent the answer data $z(t)$.
}
\label{Lorenz_MSE_vs_mu}
\end{figure*}

Our method consists of the following two steps: (I) training in the source domain with $\mathcal{D_S}$,
and (II) training in the target domain with $\mathcal{D_T}$.

(I) training in the source domain.
The readout from the ESN is given by $\hat{v}(k) = \sum_{i=1}^N w^{\mathcal{S}}_i r_i(k)$.
The hat symbol indicates the inferred quantities.
The readout weight ${\bm w}^{\mathcal{S}}$ is determined with $\mathcal{D_S}$
to minimize the mean square error (MSE)
${E}({\bm w}) = \langle ( v - \hat{v})^2 \rangle_{T_L} = \langle ( v - \sum_{i=1}^N w_i r_i )^2 \rangle_{T_L}$,
where $\langle a \rangle_T := \frac{1}{T} \sum_{k=1}^T a(k)$.
Calculating $\frac{\partial}{\partial w_j} E({\bm w}) = 0 ~ (j=1,\cdots,N)$,
the trained readout weight is given by
\begin{align}
{\bm w}^{\mathcal{S}} = R^{-1} {\bm q},
\label{RCformula}
\end{align}
where
$R_{ij} := \langle r_i r_j \rangle_{T_L}$ and $q_i := \langle v r_i \rangle_{T_L}$.

(II) training in the target domain.
We use the same ESN as in the source domain, and train the readout weight ${\bm w}^{\mathcal{T}}$.
Because of the similarity of the training data $\mathcal{D_S}$ and $\mathcal{D_T}$,
the relation ${\bm w}^{\mathcal{T}} \simeq {\bm w}^{\mathcal{S}} + \delta {\bm w}$ would hold
with a small correction $\delta {\bm w}$.
Thus, we consider the readout from the ESN as 
$\hat{v}(k) = \sum_{i=1}^N ( w^{\mathcal{S}}_i  + \delta w_i ) r_i(k)$
with the weight ${\bm w}^{\mathcal{S}}$ already trained by Eq.~(\ref{RCformula}).
The correction weight $\delta {\bm w}$ is determined 
so as to minimize the following function:
\begin{equation}
\mathcal{E}(\delta {\bm w}) = \Bigg< \Big( v'(t) - \sum_{i=1}^N (w^\mathcal{S}_i + \delta w_i) r'_i(t) \Big)^2 \Bigg>_{T'_L} + \mu \| \delta {\bm w} \|^2_2,
\end{equation}
where $\| \delta {\bm w} \|^2_2 =  \sum_{i=1}^N \delta {w}_i^2$.
The variables with primes, such as $v'$, denote variables in the target domain.
We refer to the parameter $\mu ~\in[0, \infty]$ as the {\it transfer rate}.
Calculating $\frac{\partial}{\partial \delta w_j} \mathcal{E}(\delta {\bm w}) = 0 ~ (j=1,\cdots,N)$,
the correction weight is given by
\begin{equation}
\delta {\bm w} = [R'+ \mu I]^{-1} {\bm q}',
\label{TLformula}
\end{equation}
where
$R'_{ij} := \langle r'_i r'_j \rangle_{T'_L}$, $I$ is the identity matrix and
${\bm q}' := \langle v' {\bm r}' \rangle_{T'_L} - R' {\bm w}^\mathcal{S}$.

The transfer rate $\mu$, which is similar to $l_2$ regularization,
controls the amount of knowledge transferred from the source domain to the target domain.
If the transfer rate is zero, $\mu = 0$, the above formula is reduced to the conventional RC method
which is supervised learning by using the target data $\mathcal{D_T}$ only, i.e., transferring no knowledge from the source domain.
On the other hand, in the limit of the large transfer rate, $\mu \to \infty$,
we have $\|  \delta {\bm w} \|_2 \to 0$ (see Appendix \ref{proofs} for proof).
Namely, in this limit, the above formula is reduced to a method that simply reuses the weight ${\bm w}^\mathcal{S}$,
i.e., no learning in the target domain.
The above-mentioned formula for transfer learning constitutes a one-parameter family of learning methods,
which connects the conventional RC ($\mu = 0$) and the simple transfer method ($\mu \to \infty$).

\section{Transfer learning\\
for Lorenz chaos}
To verify the effectiveness of the proposed method,
we use the Lorenz equations:
$dx/dt= \sigma (y-x),~ dy/dt= x(\rho-z)-y,~ dz/dt= xy-bz$.
We fix the parameters $\sigma, b$ to $\sigma = 10, b=8/3$
and change the parameter $\rho$ (Fig.~\ref{concept_TL}).
The task is to infer $z(t)$ from the sequence of $x(t)$ \cite{Lu2017}.
For RC, the continuous time-series $x(t)$ and $z(t)$ are sampled with a period $\tau=0.01$, and used for the input and output signal, respectively.
We assume that a sufficient amount of the training data, $\mathcal{D_S} = \{ x(t), z(t) \}_{0 \le t \le T_L}$,
is available for $\rho_{\mathcal S}=28$.
We show that our method utilizes knowledge learned from $\mathcal{D_S}$ to realize the inference in 
the target domains $\rho_{\mathcal T}= 32$ and $\rho_{\mathcal T}= 40$.

Fig.~\ref{Lorenz_MSE_vs_mu} (a) shows the MSE of the inference by the proposed method.
The source domain is  $\rho_{\mathcal S}= 28$ and the target domain is $\rho_{\mathcal T}= 32$.
To evaluate the inference accuracy by the proposed method statistically,
we perform the following training-testing procedure of the transfer learning:
\begin{algorithm}[H]
  \caption{Training-testing procedure}
   \begin{algorithmic}[1]
   \State Train ${\bm w}^{\mathcal S}$ by Eq.~(\ref{RCformula})
with $\mathcal{D_S}$
   \State Fix the transfer rate $\mu \in[0, \infty]$
   \For{$j=1,\cdots M$}
   \State Train $\delta {\bm w}^{(j)}$ by Eq.~(\ref{TLformula}) with $\mathcal{D_T}^{(j)}$
   \State Test with ${\bm w}^{\mathcal{S}} + \delta {\bm w}^{(j)}$ and output the $j$-th MSE
   \EndFor
   \end{algorithmic}
\end{algorithm}

The length of the training data $\mathcal{D_S}$ in the source domain is $T_L = 10^4$.
In the above procedure, $M$ denotes the number of samples in the training data, $\mathcal{D_T}^{(j)} = \{ x^j(t), z^j(t)  \}_{0 \le t \le T'_L} ~~(j=1,\cdots,M)$, in the target domain.
We use $M=100$ and $T'_L = 5~ (\ll T_L)$.
Each training data sample only includes less than ten cycles around the fixed points of the Lorenz attractor.
Figs.~\ref{concept_TL} (c) and (c') present examples of the training data.
The MSE is the generalization error with common test data in the target domain,
which differ from the training data,
with the length $T_{\text{test}} = 5 \times 10^3$.
The median MSE over $M=100$ samples of training data, indicated by the red circle,
is plotted for each value of $\mu$.
The blue shaded area indicates the range of the MSE from the first to the third quartile, which characterizes the statistical dispersion of the MSE.
In the inset,
we show the same result for the MSE, plotted using the logarithmic values,
and the dashed (solid) line represents the median MSE in the case of $\mu=0$ ($\mu=\infty$).
At each end, the proposed method is reduced to the conventional and simple transfer methods, respectively.
In the case of $\mu = \infty$, we set $\delta {\bm w}={\bm 0}$.

\begin{figure}[hb]
\includegraphics[width=8.6cm]{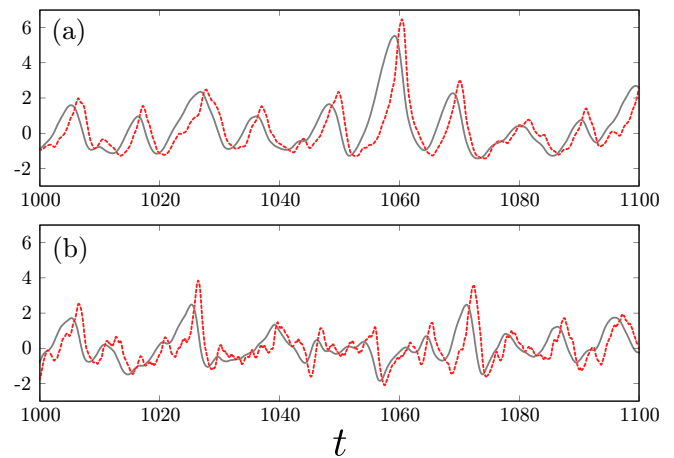}
\caption{Normalized time-series of the energy $K(t)$ and energy dissipation rate $\epsilon(t)$ at (a) $R_\lambda \simeq 35$ and (b) $R_\lambda \simeq 120$.
The gray solid line and red broken line represent $K(t)$ and $\epsilon(t)$, respectively.}
\label{hit_ts}
\end{figure}

In the case of $\rho_{\mathcal T} = 32$, the attractor is expected to be similar to that at $\rho_{\mathcal S}=28$,
and thus, transfer learning is effective.
In fact, as shown in Fig.~\ref{Lorenz_MSE_vs_mu} (a), 
if we conduct the transfer learning with the optimal transfer rate $\mu \simeq 10^{-8}$,
the median of the MSEs decreases drastically
and it becomes smaller than, by an order of magnitude,
that of the conventional method ($\mu=0$).
Note that the statistical dispersions are also reduced.

In practice, we must find the optimal parameter with a small amount of data.
Even in such a situation,
i.e., $M$ is small and the length of the test data $T_{\text{test}}$ is short,
the above procedure gives an estimation of the optimal transfer rate as will be discussed in \S \ref{optimization}.

\begin{figure*}[ht]
\includegraphics[width=17.2cm]{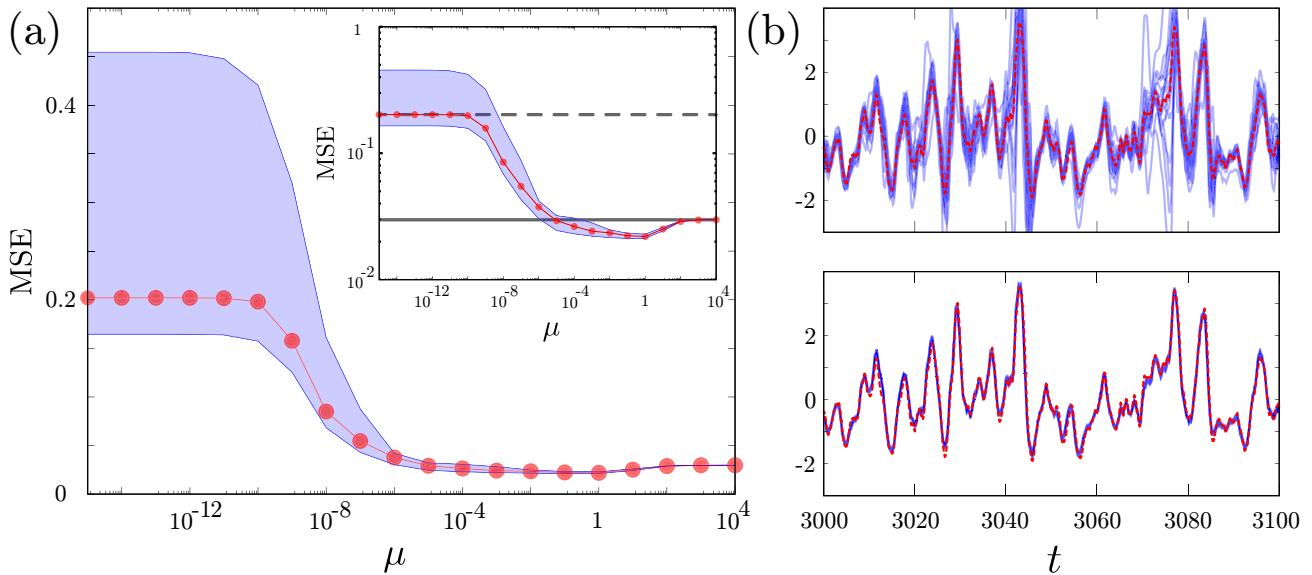}
\caption{(a) Generalization error
(mean square error, MSE) as a function of the transfer rate $\mu$.
Similar plots to Fig.~3 but for the inference of the energy dissipation rate of
the Navier-Stokes turbulence at $R_\lambda \simeq 120$.
The red circles represent the median MSE. The blue shaded area indicates the range of MSEs from the first to the third quartile.
Inset: the logarithmic graph of the same MSE data.
The broken and solid lines represent the median MSE in the cases of $\mu=0$ and $\mu=\infty$, respectively.
(b) The blue lines are the inferred values 
$\hat{\epsilon}^{(j)}(t)~~ (j=1,\cdots, 20)$ at $R_\lambda \simeq 120$
by the conventional method ($\mu=0$, upper panel) and the proposed transfer learning ($\mu=1$, lower panel).
The red broken lines represent the answer data $\epsilon (t)$.
}
\label{hit_MSE}
\end{figure*}

The time-series of the values $\hat{z}(t)$ inferred by the conventional method
and transfer learning at the optimal transfer rate
are shown in the upper and lower panels in Fig.~\ref{Lorenz_MSE_vs_mu} (b), respectively.
In the figure, 
the solid blue lines of the inferred $\hat{z}^{(j)}(t)$ correspond to the each training data sample, $\mathcal{D_T}^{(j)} ~ (j=1,\cdots,100)$,
and the broken red line represents the answer data $z(t)$.
When we use the conventional method,
the inferred time-series significantly differs from the true time-series.
On the other hand, the transfer learning method enables
the time-series of $z(t)$ to be inferred with a smaller error and less statistical dispersion.

Even in the case of $\rho_{\mathcal T} = 40$, which is far from the source domain,
transfer learning can reduce the median MSE and the statistical dispersion
compared with the conventional method (see Appendix \ref{lorenz_28to40}).

\section{Transfer learning\\
for Fluid Turbulence}
We tackle a critical problem associated with turbulence: the inference of the energy dissipation rate $\epsilon(t)$.
Although the energy dissipation rate plays an important role in statistical turbulence theory and turbulence modeling,
the direct measurement of $\epsilon(t)$ is difficult.
As an easily measurable quantity,
we use the kinetic energy $K(t)$ of the turbulent flow.
The task is to infer $\epsilon(t)$ from $K(t)$.

As described before, the ML-based turbulence modeling requires training data of high-Reynolds-number turbulence,
such as $\{ K(t), \epsilon(t) \}$,
collected over a long period of time.
However, such turbulence data are not available in practice.
On the other hand, the calculation of turbulence at {\it low} Reynolds number is much easier.
Since the energy cascade dynamics is insensitive to variation in the Reynolds numbers (see \cite{goto_inubushi} for the details),
we expect that there is a similarity of turbulence attractors over a wide range of the Reynolds numbers.
Hence, the knowledge learned from turbulent data at a low Reynolds number
to be useful for the same task at a high Reynolds number.
This gives a reason why transfer learning is effective.

We conducted direct numerical simulations of the Navier-Stokes equations with a steady forcing term in a periodic box,
using the Fourier spectral method.
For the temporal integration,
the fourth-order Runge--Kutta--Gill scheme was used (see Appendix \ref{dns} for details).

Here we use the time-series of the spatial average of the energy and the energy dissipation rate at $R_\lambda \simeq 35$
as the training data $\mathcal{D_S} = \{ K(t), \epsilon(t)\}_{0 \le t \le T_L}$ in the source domain,
and those at $R_\lambda \simeq 120$ as the training data $\mathcal{D_T} = \{ K(t), \epsilon(t)\}_{0 \le t \le T'_L}$ in the target domain,
where $R_\lambda$ is the Taylor micro-scale Reynolds number
and it plays the role of $\rho$.
We emphasize that the integral-scale Reynolds number $R$ in the target domain is approximately an order of magnitude higher than that in the source domain, since $R \propto R_\lambda^2$ \cite{frisch1995}.

Fig.~\ref{hit_ts} shows the normalized time-series of the energy $K(t)$ and energy dissipation rate $\epsilon(t)$ at (a) $R_\lambda \simeq 35$ and (b) $R_\lambda \simeq 120$.
The gray solid line and red broken line represent $K(t)$ and $\epsilon(t)$, respectively.
The mean turnover time of the largest eddies, defined by $\braket{T} = \braket{L/\sqrt{2K/3}}$
with $L$ being the integral length,
are $\braket{T} \simeq 0.7$ at $R_\lambda \simeq 35$ and $\braket{T} \simeq 0.5$ at $R_\lambda \simeq 120$.
The energy dissipation rate $\epsilon(t)$ at $R_\lambda \simeq 35$ changes in time following the energy $K(t)$ with a delay
owing to the energy cascade.
At $R_\lambda \simeq 120$, the time delay is still found in the relation between $K(t)$ and $\epsilon(t)$;
however, the relation becomes more than merely a delay.

Training data collected for a sufficiently long time $\mathcal{D_S}$ with $T_L=3.0 \times 10^3$ are used to obtain ${\bm w}^\mathcal{S}$ at $R_\lambda \simeq 35$.
We assume that the amount of available training data is highly limited for the calculation of $\delta {\bm w}$ at the target domain, $R_\lambda \simeq 120$.
The length of the each training data of $\mathcal{D_T}^{(j)} ~ (j=1,\cdots,M)$ is $T'_L = 50 ~(\ll T_L)$,
which includes roughly ten quasi-periodic cycles of the energy cascade events \cite{goto_2016}.
The number of samples of training data is $M=20$.
The length of the test data is $T_{\text{test}}=2.0 \times 10^3$.
For RC, the continuous time-series $K(t)$ and $\epsilon(t)$ are sampled with a period $\tau = 0.2$, and used for the input and output signal, respectively.

Fig.~\ref{hit_MSE} (a),
which uses the same symbols and lines as in Fig.~\ref{Lorenz_MSE_vs_mu},
shows the dependency of the inference MSE on the transfer rate $\mu$.
The most accurate inference (the smallest MSE) is achieved by the proposed transfer learning method with $\mu \simeq 1$,
which implies that
the ESN learned from the low-Reynolds-number turbulence data
requires only slight corrections by the high-Reynolds-number turbulence data.
Compared with the conventional method ($\mu=0$),
transfer learning effectively reduces the inference errors and the statistical dispersion.
As mentioned above, there is the similarity between the training data $\mathcal{D_S}$ and $\mathcal{D_T}$ for this particular task,
which is explained by the energy cascade dynamics of turbulence \cite{goto_inubushi}.


The corresponding time-series of the inferred value $\hat{\epsilon}(t)$ in the target domain, $R_\lambda \simeq 120$,
by the conventional method and transfer learning with the optimal transfer rate ($\mu=1$)
are shown in the upper and lower panels in Fig.~\ref{hit_MSE} (b), respectively.
In each panel, the solid blue lines represent the inferred $\hat{\epsilon}^{(j)}(t)  ~ (j=1,\cdots,M)$,
corresponding to each training data sample $\mathcal{D_T}^{(j)}$,
and the broken red line represents the answer time-series $\epsilon(t)$.

The conventional method produces a large inference error and considerable statistical dispersion.
On the other hand, the transfer learning method
achieves almost perfect inference of the energy dissipation rate.

\section{Estimation of optimal transfer rate with a small amount of data}
\label{optimization}

In the previous sections, we used a large amount of data, e.g., a large number of samples of training data and a sufficiently long test data, in order to statistically verify the performance of the transfer learning method.
However, in practice, we need to estimate the optimal transfer rate from a small amount of data.

In order to investigate the optimization with a small amount of data,
we conduct numerical experiments in both cases of the Lorenz equations and the Navier-Stokes equations.
The target domain of the Lorenz case is $\rho_{\mathcal{T}}=32$.
Here 
we assume that we can use a single sample of data, i.e., $M=1$, in the target domain with the length of $3 T'_L$,
and divide it into three.
The first one ($0 \le t < T'_L$) is used to obtain the reservoir state being synchronized with the input signal,
the second one ($T'_L \le t < 2 T'_L$) is used for training,
and the third one ($2T'_L \le t < 3 T'_L$) is used for testing.
The other settings of the experiments including the values of $T'_L$ for the Lorenz and Navier-Stokes equations are the same as in the previous sections.

\begin{figure*}[t]
\includegraphics[width=15cm]{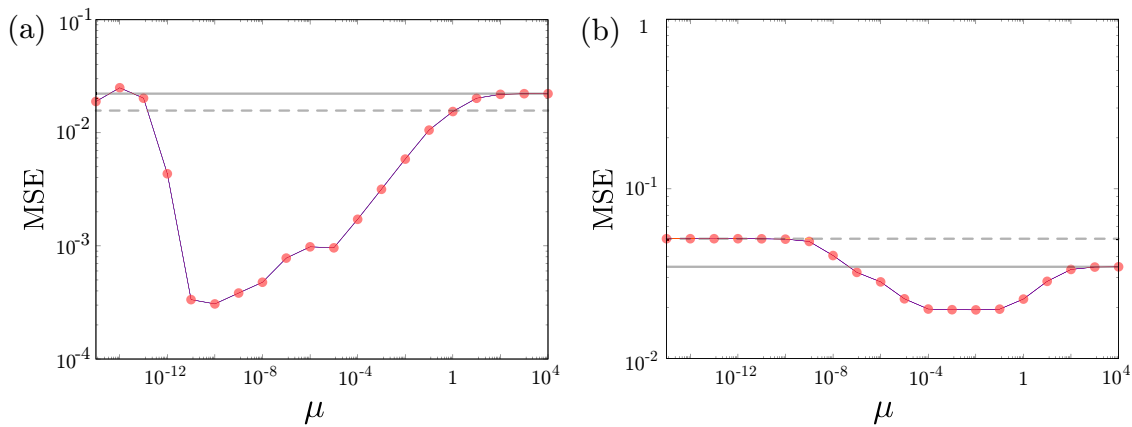}
\caption{
MSE (red circles) of the inference task for (a) the Lorenz equations
and (b) the Navier-Stokes equations,
as a function of the transfer rate $\mu$.
The target domain of the Lorenz case is $\rho_{\mathcal{T}}=32$.
The MSE is estimated with a short test data $T_{\text{test}} = T'_L$ and $M=1$.
The broken and solid lines represent the MSE in the cases of $\mu=0$ and $\mu=\infty$, respectively.}
\label{opt}
\end{figure*}

Figures~\ref{opt}(a) and (b) show the MSE of the inference task for the Lorenz equations
and for the Navier-Stokes equations, respectively.
For the Lorenz equations,
while the inset of Fig.~2(a) shows the optimal transfer rate is in the range $10^{-9} \le \mu \le 10^{-6}$,
Fig.~\ref{opt}(a) shows it is in the range $10^{-11} \le \mu \le 10^{-8}$.
For the Navier-Stokes equations,
while the inset of Fig.~4(a) shows the optimal transfer rate is in the range $10^{-2} \le \mu \le 1$,
Fig.~\ref{opt}(b) shows it is in the range $10^{-4} \le \mu \le 10^{-1}$.
These demonstrations suggest that we can estimate the optimal transfer rate from such a small amount of data,
in particular, for the turbulence case,
the data for at most $15$-fold of the correlation time \cite{Goto2016} is sufficient for the optimization.


\color{black}
\section{Conclusions}
We have developed the transfer learning method of RC,
and shown that the optimization of the transfer rate is essential 
for the inference problem of the Lorenz chaos.
Furthermore, 
if we choose a suitable physical quantity for learning, as shown in the successful inference of the energy dissipation rate of turbulence,
the amount of learning in the target domain can be drastically reduced.

In this paper, we showed in a statistically reliable manner
that there exists an optimal transfer rate
by using a large amount of data.
%
However, in practice,
we must find an optimal transfer rate with a small amount of data, i.e.,
$M$ is small and the length of the test data $T_{\text{test}}$ is short.
As demonstrated in \S \ref{optimization},
the transfer learning method gives an estimation of the optimal transfer rate
even when only a small amount of data is available.
To improve the accuracy of the estimation,
it will be useful to employ ML-techniques such as the cross-validation
or the quantification of 
similarity between attractors in the source and target domains.

We implemented transfer learning for parameter changes of the dynamical systems.
Once we train a reservoir computer for a dynamical system with a certain parameter,
the proposed method can eliminate most of the computational cost of training for the same dynamical system with different parameters.
The applications of the proposed method are not restricted to the parameter change;
for instance, it is possible to train a reservoir computer with numerical simulations,
and then utilize it in predictions for physical experiments via the proposed transfer learning.
Although the present study focused on the nonlinear dynamics,
transfer learning for RC is a model-free flexible ML-method, and hence, can be applied to any other physical system.
Physical RC, physical implementations of RC using physical devices such as lasers,
is highly active research topic \cite{Sunada2019, PhysRevApplied.12.034058, Tanaka2019},
and the transfer learning is also useful to realize the physical RC.

We hope that the proposed method
will open up new possibilities of ML methods for nonlinear dynamics, and
will be an effective tool for the long-standing problems in physics.
In particular, 
we expect 
the present study
to be a crucial step toward the development of the ML-based turbulence modeling that
utilizes the attractor similarity associated with the universality of the small-scale statistics of turbulence.

\appendix
\section{Required length of training data\\
for echo state network}
\label{data_length}

Usually, machine learning methods such as RC require a large amount of training data.
In this appendix, firstly,
we estimate the required amount of training data explicitly in the general setting of the {\it linear} ESN.
Then, focusing on the turbulence case studied in the main text,
we discuss the required length of the training data, and describe the numerical results for the nonlinear ESN.

Training of the ESN requires
the convergence of $\braket{ r_i r_j }_{t}$ and $\braket{r_i v}_{t}$,
where $\langle a \rangle_t := \frac{1}{t} \sum_{k=1}^t a(k)$.
For the linear ESN,
we can generally estimate the required length of training data.
The linear ESN is a signal-driven dynamical system:
${\bm r} (k+1) = J {\bm r} (k) + \epsilon u(k) {\bm 1}$,
where all the components of the vector ${\bm 1}$ are one.
We obtain the following expression
\begin{align}
{\bm r} (k+1) 
&= J {\bm r} (k) + \epsilon u(k) {\bm 1} \nonumber\\
&= J^m {\bm r} (k-m+1) + \epsilon \sum_{\ell=0}^{m-1} u(k-\ell) J^\ell {\bm 1} \nonumber\\
&\to \epsilon \sum_{\ell=0}^{\infty} u(k-\ell) J^\ell {\bm 1} ~~(m \to \infty).
\end{align}
In the last step, we assume
the spectral radius of the matrix $\rho(J)$ is strictly less than one, $\rho(J) < 1$,
which ensures synchronization with the input signal.
Therefore, we have
\begin{align}
\braket{ r_i r_j }_{t} = \epsilon^2 \sum_{\ell, \ell'=0}^\infty [J^\ell {\bm 1}]_i [J^{\ell'} {\bm 1}]_j C_{uu}(\ell-\ell'),
\end{align}
and 
\begin{align}
\braket{ r_i v }_{t}
&= \epsilon \sum_{\ell=0}^\infty [J^\ell {\bm 1}]_i C_{uv}(\ell),
\end{align}
where $C_{uu}(\ell-\ell')$ is the auto-correlation function,
$C_{uu}(\ell-\ell') := \braket{ u(k-\ell) u(k-\ell')}_{t}$,
and 
$C_{uv}(\ell)$ is the cross-correlation function,
$C_{v}(\ell) := \braket{ u(k-\ell) v(k)}_{t}$.
Thus,
the convergence of $\braket{ r_i r_j }_{t}$ and $\braket{r_i v}_t$ requires the convergence of $C_{uu}(\ell-\ell')$ and $C_{uv}(\ell)$, respectively.

Considering the inference task of the energy dissipation rate as discussed in the main text,
$u = K$ and $v = \epsilon$,
and sufficient training data is necessary
such that the auto-correlation $C_{KK}$ and the cross-correlation $C_{K \epsilon}$ converge.
The length of such the data
exceeds the turnover time of the largest eddies by several thousand times \cite{Goto2016}.
For the nonlinear ESN used in this paper,
we have numerically studied the convergence of $\braket{r_i r_j}_t$ and $\braket{r_i v}_t$,
and confirmed that the convergence of these values requires a long period of turbulence data
as mentioned above.

%

\section{Proofs of asymptotic formulas\\
of the transfer learning methods}
\label{proofs}

\subsection{Reduction to conventional RC ($\mu=0$)}

When $\mu=0$, Eq. (3) of the main text becomes
\begin{align}
\delta {\bm w} &= R'^{-1} {\bm q}' \nonumber \\
&= R'^{-1} \Big( \Braket{ v' {\bm r}'}_{T'} - C' {\bm w}^{\mathcal{S}} \Big) \nonumber\\
&= R'^{-1}  \Braket{ v' {\bm r}'}_{T'}  - {\bm w}^{\mathcal{S}}.
\end{align}
Thus, ${\bm w}^{\mathcal{S}}+ \delta {\bm w}= R'^{-1} \Braket{ v' {\bm r}'}_{T'}$, which is
Eq. (1) in the main text for the conventional RC.\\

\subsection{Simple transfer method ($\mu \to \infty$)}

In the limit of $\mu \to \infty$, we have
\begin{align}
\| \delta {\bm w} \| &= \| [R' + \mu I]^{-1}  {\bm q}' \| \nonumber \\
&\le \| [R' + \mu I]^{-1}  \| \cdot \|  {\bm q}' \| \nonumber \\
&= \frac{1}{\mu} \| (I - T_\mu)^{-1} \| \cdot \|  {\bm q}' \| ~~ (T_\mu := - R'/\mu) \nonumber \\
&\le  \frac{ \|  {\bm q}' \|}{\mu (1- \| T_\mu \|)} \nonumber\\
&\to 0 ~~(\mu \to \infty).
\end{align}
In the last step, we have used the Neumann series.

\section{Transfer learning for Lorenz chaos\\
far from source domain}
\label{lorenz_28to40}

\begin{figure}[t]
\includegraphics[width=8.6cm]{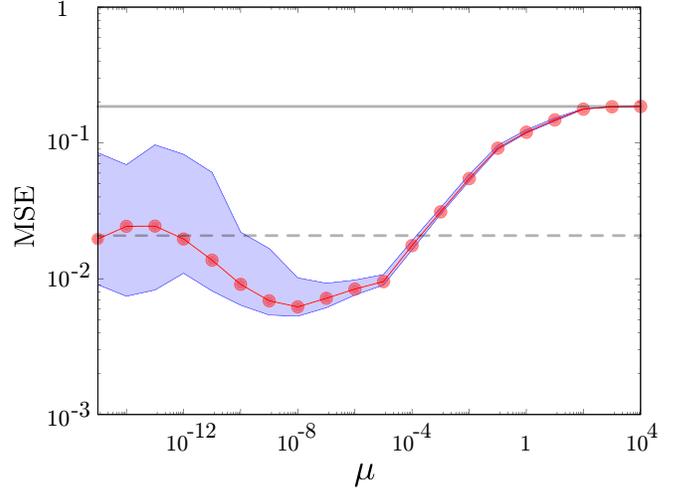}
\caption{MSE of the inference task for the Lorenz equations with $\rho_{\mathcal{T}}=40$
as a function of the transfer rate $\mu$.
The red circles represent
the median MSE.
The blue shade indicates the range in MSEs from the first to the third quartile.
The broken and solid lines represent the median MSE in the cases of $\mu=0$ and $\mu=\infty$, respectively.}
\label{MSE28to40}
\end{figure}

Even in the case of $\rho_{\mathcal T}= 40$, which is far from the source domain,
the transfer learning can reduce the median MSE and the statistical dispersion
compared with the conventional method.
The results are presented in Fig.~\ref{MSE28to40}, in which
the symbols and lines have the same meaning as those in the main text.

\begin{figure}[t]
\includegraphics[width=8.6cm]{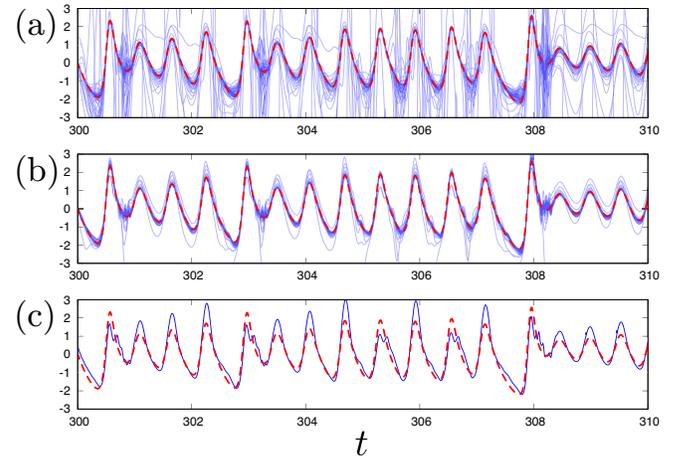}
\caption{The time-series of the inferred values $\hat{z}^{(j)}(t)~~ (j=1,\cdots,M)$ at $\rho_{\mathcal{T}} = 40$ corresponding to the $M=100$ samples of the training data. 
The inferred values
by the conventional ($\mu=0$), the transfer learning ($\mu=10^{-8}$) and the simple transfer method ($\mu=\infty$)
are shown in (a), (b), and (c), respectively.
The red broken lines depict the answer data $z(t)$.}
\label{ts_28to40}
\end{figure}

Compared with the case of $\rho_{\mathcal T}= 32$,
the similarity between the attractors in the source domain and the target domain are lower.
In fact, the simple transfer method ($\mu=\infty$) is ineffective.
Although the target domain is far from the source domain,
the transfer learning with the optimal transfer rate $\mu \simeq 10^{-8}$ still reduces
the median MSE and the statistical dispersions
compared with the conventional method ($\mu=0$).

The time-series of the inferred values $\hat{z}(t)$ are shown in Fig. \ref{ts_28to40}.
Panels (a), (b), and (c) correspond to the results with the conventional ($\mu=0$),
the transfer learning ($\mu=10^{-8}$), and the simple transfer method ($\mu=\infty$), respectively.
%
%
The conventional method leads to large errors and statistical dispersions.
When the simple transfer method is used,
inference errors around the maximal values of $z(t)$ inevitably arise.
On the other hand, when the transfer learning method is used,
the time-series of $z(t)$ can be inferred with a smaller error and less statistical dispersion.

\section{Direct numerical simulations\\
of the Navier-Stokes equations}
\label{dns}

We numerically solve the three-dimensional Navier-Stokes equations
in a periodic box, using the Fourier spectral method.
The aliasing errors are removed by the phase shift method.
In particular, the vorticity equation,
\begin{align}
\frac{\partial {\bm \omega}}{\partial t} = {\bm \nabla} \times ( {\bm u} \times {\bm \omega})
+ \nu \nabla^2 {\bm \omega} + {\bm \nabla} \times {\bm f},
\end{align}
is integrated in time with the fourth-order Runge-Kutta-Gill scheme,
where $\bm \omega = {\bm \nabla} \times {\bm u}$.
The forcing term \cite{Goto2017} is
\begin{align}
{\bm f}(x,y,z) =
\left[
    \begin{array}{c}
      - \sin (2 \pi x /{\mathcal{L}}) \cos (2 \pi y /{\mathcal{L}}) \\
      +  \cos (2 \pi x /{\mathcal{L}}) \sin (2 \pi y /{\mathcal{L}}) \\
      0
    \end{array}
  \right],
\end{align}
where $\mathcal{L}$ is the length of the side of the period box.
The parameters of
the number $n^3$ of the Fourier modes,
the kinematic viscosity $\nu$ of the fluid,
the step $\Delta t$ for the temporal integration,
the mean of the Taylor micro-scale Reynolds number $\braket{R_{\lambda}}$,
and the mean turnover time of the largest eddies
$\braket{T}$
are summarized in TABLE \ref{table}.

\begin{table}[h]
 \caption{Parameters and statistics.}
 \label{table}
 \centering
  \begin{tabular}{cccccc}
   \hline
   Domain & $~n^3~~$ & $\nu~~$ & $\Delta t~~$ & $\braket{R_{\lambda}}~~$ & $\braket{T}$ \\
   \hline \hline
   Source & $~{32}^3~~$ & $0.064~~$ & $4 \times 10^{-3}~~$ & $35~~$ & $0.7$ \\
   \hline
   Target & $~{128}^3~~$ & $0.008~~$ & $2 \times 10^{-3}~~$ & $120~~$  & $0.5$\\
   \hline
  \end{tabular}
\end{table}

\color{black}

\begin{acknowledgments}
This work was partly supported by
JSPS 
Grant-in-Aid for Early-Career Scientists No.~19K14591,
JSPS Grants-in-Aid for Scientific Research No.~16H04268, and
The Nakajima Foundation.
The direct numerical simulations of the Navier-Stokes equations were conducted using the supercomputer systems of the Japan Aerospace Exploration Agency (JAXA-JSS2).
\end{acknowledgments}

\nocite{*}

\bibliography{TLRC}

\end{document}